\documentclass[
 reprint,
 amsmath,amssymb,
 aps,
 prl,
 floatfix,
]{revtex4-2}

\usepackage{graphicx}
\usepackage{dcolumn}
\usepackage{bm}
\usepackage[colorlinks=true, allcolors=blue]{hyperref}
\usepackage{lipsum}
\usepackage{braket}
\usepackage{booktabs}
\usepackage{multirow}
\usepackage{siunitx}

\begin{document}


\title{Complementing the transmon by integrating a geometric shunt inductor}

\author{Simone D. Fasciati}
\email[Corresponding author: ]{simone.fasciati@physics.ox.ac.uk}
\author{Boris Shteynas}
\altaffiliation{Present address: Oxford Quantum Circuits, Thames Valley Science Park, Shinfield, Reading RG2 9LH, UK}
\author{Giulio Campanaro}
\altaffiliation{Present address: Alice \& Bob, 49 Bd du Général Martial Valin, 75015 Paris, France}
\author{Mustafa Bakr}

\author{Shuxiang Cao}
\author{Vivek Chidambaram}
\altaffiliation{Present address: National Quantum Computing Centre, Harwell Campus, Didcot OX11 0QX, UK}
\author{James Wills}
\altaffiliation{Present address: Oxford Quantum Circuits, Thames Valley Science Park, Shinfield, Reading RG2 9LH, UK}
\author{Peter J. Leek}
\email[Corresponding author: ]{peter.leek@physics.ox.ac.uk}
\affiliation{Department of Physics, Clarendon Laboratory, University of Oxford, Oxford OX1 3PU, UK}

\date{\today}

\begin{abstract}

We realize a single-Josephson-junction transmon qubit shunted by a simple geometric inductor. We couple it capacitively to a conventional transmon and show that the ZZ interaction between the two qubits is completely suppressed when they are flux-biased to have opposite-sign anharmonicities. Away from the flux sweet spot of the inductively-shunted transmon, we demonstrate fast two-qubit interactions using first-order sideband transitions. The simplicity of this two-qubit-species circuit makes it promising for building large lattices of superconducting qubits with low coherent error and a rich gate set.

\end{abstract}

\maketitle


Several hardware platforms have shown the potential for building a quantum computer that can tackle challenging computational problems by encoding and processing information according to the laws of quantum mechanics~\cite{fedorov_quantum_2022, ion, neutral, RevModPhys.95.025003}. Superconducting quantum circuits are among the leading technologies, with circuits now realized at the scale of hundreds of qubits~\cite{arute_quantum_2019, kim_evidence_2023} and basic quantum error correction demonstrated~\cite{zhao_realization_2022, krinner_realizing_2022, acharya_suppressing_2023}. Further reduction of errors is essential for the realization of useful quantum computing.

The current workhorse of superconducting circuits is the transmon qubit~\cite{Koch2007}, a simple and reliable circuit achieved by shunting a Josephson junction (JJ) with a capacitor. This realizes a non-linear oscillator with eigenstates insensitive to charge noise, at the cost of a modest anharmonicity equal to only a few \% of its transition frequency. This limits the attainable speed of single-qubit gates~\cite{chow2010Optimized, werninghaus_leakage_2021}, as well as the performance of two-qubit operations due to the interaction with nearby non-computational levels, causing an always-on dispersive interaction in the computational subspace~\cite{Krinner2020Benchmarking, zhao2022Crosstalk}. The latter is typically quantified by the static cross-Kerr (or ZZ) shift

\begin{equation}
\label{eq:zeta_def}
    \zeta = E_{\ket{\widetilde{11}}}-E_{\ket{\widetilde{10}}} - E_{{\ket{\widetilde{01}}}}+E_{\ket{\widetilde{00}}},
\end{equation}
where $E_{\ket{\widetilde{ij}}}$ denotes the energy of the eigenstate $\ket{\widetilde{ij}}$ in the dressed (coupled) two-qubit system. The labels $i,j$ refer to the number of excitations present in either of the qubits. The ZZ shift $\zeta$ is a measure of how much the frequency of one qubit depends on the state of the other, and causes coherent error in a quantum computation. 

Several strategies have been explored to mitigate this error source, e.g. by using a higher-anharmonicity qubit such as the fluxonium~\cite{manucharyan_fluxonium_2009, bao22fluxonium, moskalenko_high_2022}, introducing additional modes and coupling paths to achieve $\zeta=0$ via interference~\cite{Houck2019, IBM2021, Sung2021, IQM, li2024realization}, or actively driving the system to offset the static ZZ shift~\cite{Sizzle, noguchi_fast_2020, mitchell_hardware-efficient_2021}. 

An alternative solution can be derived by considering a simplified picture of two Duffing oscillators with small anharmonicities $\alpha_1,\alpha_2$ coupled via a static exchange interaction $J$. The ZZ shift can in this case be approximated as $\zeta \approx 2J^2(\alpha_1 + \alpha_2)/\Delta^2$, with $\Delta$ the qubit-qubit detuning~\cite{zhao_high-contrast_2020}. This expression reduces to zero for opposite anharmonicities, $\alpha_1 = -\alpha_2$. While this requires the use of two distinct species of qubit, it has the advantage of completely passive ZZ suppression with no additional coupling hardware required. This approach has been previously described theoretically~\cite{zhao_high-contrast_2020, xu_zz_2021}, and realized experimentally in a two-qubit system consisting of a transmon and a capacitively shunted flux qubit (CSFQ)~\cite{ku_suppression_2020-1}. 

The CSFQ has a plasmonic excitation spectrum similar to the transmon, but with a positive anharmonicity at the half-flux bias point~\cite{FluxQubitRevisited}. It approximately realizes a radio-frequency superconducting quantum interference device (RF-SQUID), i.e. a closed loop formed by a JJ shunted with a linear inductor, described by the Hamiltonian 
\begin{equation}
\label{eq:RF-SQUID_Hamiltonian}
    \mathcal{H}_{RF} = 4E_C\hat{n}^2-E_J\cos(\hat{\phi}+\phi_e)+\frac{1}{2}E_L\hat{\phi}^2,
\end{equation}
where $E_C$ is the total charging energy, $E_J$ is the Josephson energy, and $E_L$ is the linear inductive energy. The variable $\phi_e=2\pi\Phi/\Phi_0$ represents an external magnetic flux bias $\Phi$ threading the RF-SQUID, with $\Phi_0$ the magnetic flux quantum. While the CSFQ employs a small array of JJs to mimic the linear term $E_L$, one can directly use a linear geometric inductor to achieve the same effect. When the linear inductance is small, i.e. $E_L>E_J$,  the phase potential $U(\hat{\phi})= -E_J\cos(\hat{\phi}+\phi_e)+E_L\hat{\phi}^2/2$ takes a single minimum for any value of $\phi_e$ due to the strong parabolic confinement of the linear inductor. This realizes a single-well plasmonic spectrum with positive anharmonicity $\alpha>0$ at $\phi_e=0.5$ (see Supplemental Material~\cite{SupplementalMaterial} for details, which includes additional Refs.~\cite{Yan2020, UnpublishedMUXpaper, Unpublishedfluxpaper, yoshihara_decoherence_2006, koch_model_2007, kumar_origin_2016, braumuller_characterizing_2020, bialczak_1f_2007}).

In this work, we experimentally realize an RF-SQUID qubit with a single JJ and a geometric meandering inductor with $E_L \gtrsim E_J$. The simplicity of this circuit, which we refer to as an \textit{inductively shunted transmon} (IST), makes it an attractive solution for enriching the physics of existing transmon-based circuits with minimal added complexity. Other realizations of RF-SQUID qubits with linear inductors have been previously demonstrated at the single-qubit level~\cite{peruzzo_geometric_2021, hyyppa_unimon_2022, Kinemon2024, hassani_inductively_2023}. Here we implement the IST in a compact, tileable, 3D-integrated design~\cite{Rahamim, Spring2022}. We first characterize its basic properties and then explore the two-qubit dynamics of a coupled transmon-IST system, showing both ZZ suppression as well as a wide range of novel qubit-qubit interactions that can be harnessed to perform entangling gates.

\begin{figure}[t!]
\includegraphics[width=8.5cm]{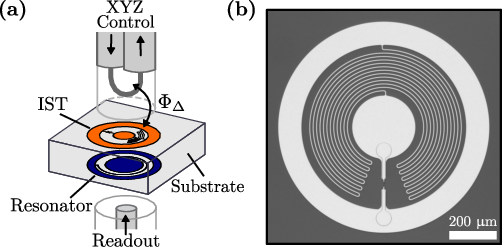}
\caption{\label{fig:IST_Basics} (a) Schematic drawing of a gradiometric inductively shunted transmon (IST) embedded in a coaxial circuit QED architecture, with off-chip control and readout. The XYZ control line can apply a differential flux $\Phi_\Delta$ to the gradiometric SQUID loop. (b) Optical micrograph of an individual IST, with identical design to device A. Aluminum metal is visible in light gray, silicon substrate in dark grey.}
\end{figure}

Our experimental design is shown schematically in Fig.~\ref{fig:IST_Basics}~(a). An optical image of a single IST can be found in Fig.~\ref{fig:IST_Basics}~(b). The IST and the coaxial lumped resonator used for dispersive readout are fabricated from aluminum thin films, located on opposing sides of a silicon substrate, and packaged inside a micro-machined aluminum enclosure.  Microwave control (XY) and flux control (Z) of the IST are provided by an off-chip differential XYZ line that offers well-matched control with low crosstalk~\cite{Unpublishedfluxpaper}. We choose a gradiometric topology for the RF-SQUID loop~\cite{braumuller_concentric_2016, gusenkova_operating_2022}, meaning that the device is sensitive to the difference $\Phi_\Delta = \Phi_1 - \Phi_2$ of the flux values through the two halves of the loop. Readout tones are applied to the resonator from a separate coaxial port. The IST inductor has a width of 5~\unit{\micro\meter} and is patterned in the same lithographic step that defines the large electrodes. More details on device fabrication can be found in the Supplemental Material~\cite{SupplementalMaterial}.

\begin{figure}[b!]
\includegraphics[width=8cm]{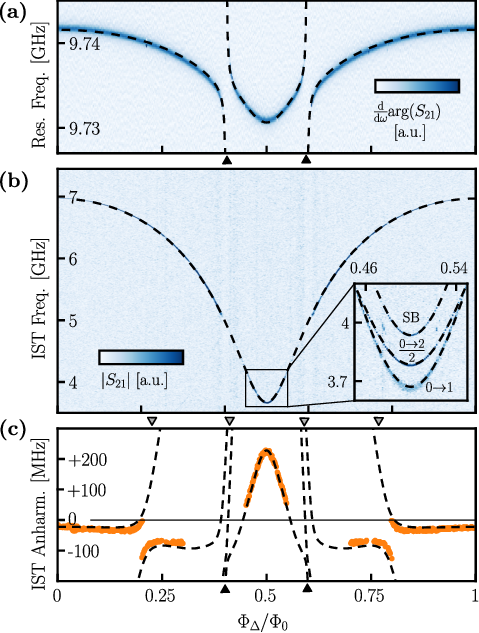}
\caption{\label{fig:IST_Spectroscopy} (a)--(b) Low-power spectroscopy of the readout resonator and IST, respectively, as a function of differential flux $\Phi_\Delta$ applied to the IST. Inset in (b) shows high power spectroscopy close to the half-flux point, revealing the two-photon transition $(0\rightarrow2)/2$ and a higher-order sideband (SB) transition~\cite{SupplementalMaterial}. (c) Flux-dependent anharmonicity of the IST obtained from high-power spectroscopy. Simultaneous numerical fit to the datasets in (a)--(c) is performed to extract circuit parameters (dashed lines). Black (gray) triangles denote the locations of avoided crossings between the IST $0\rightarrow2$ transition and the readout mode (parasitic inductor modes).}
\end{figure}

Spectroscopic measurements of an individual IST, referred to as device A, are shown in Fig~\ref{fig:IST_Spectroscopy}. The experiments are carried out in a cryogenic setup at 20~mK~\cite{SupplementalMaterial}. Continuous-wave spectroscopy of the readout resonator as a function of applied differential flux $\Phi_\Delta$ in Fig.~\ref{fig:IST_Spectroscopy}~(a) shows a characteristic periodic behavior of the resonator frequency due to dispersive coupling to the IST, which we use to calibrate the x-axis in units of $\Phi_0$ (modulo integer multiples). Tracking the readout frequency at each flux bias, we then perform spectroscopy on the IST by applying a drive tone to the XYZ line, see Fig.~\ref{fig:IST_Spectroscopy}~(b). The fundamental mode of the IST, denoted by $0\rightarrow 1$, is found at 6.98~GHz at integer flux quanta and decreases to 3.67~GHz at the half-flux bias point $\Phi_\Delta = 0.5 \Phi_0$. At low drive powers, the spectrum is clean and the transition does not show any significant avoided crossings. At higher drive powers, more transitions become visible, as shown in the inset of Fig.~\ref{fig:IST_Spectroscopy}~(b). We identify the feature just above the fundamental transition as the two-photon transition $[0\rightarrow 2]/2$ involving the second IST level. From this we can directly extract the anharmonicity $\alpha_{\mathrm{IST}}$, plotted in Fig.~\ref{fig:IST_Spectroscopy}~(c), showing that it is highly tunable with flux. The maximum value $\alpha_{\mathrm{IST}}^{max}=+228$~MHz is achieved at the half-flux point, while moving towards integer bias causes it to rapidly decrease and become negative.

The obtained spectra, including for the readout resonator, are numerically fitted to extract circuit parameters, with the result overlaid on the experimental data in Fig.~\ref{fig:IST_Spectroscopy}. For an accurate fit, it is necessary to include self-resonances of the large geometric inductor (parasitic modes), which significantly distort the IST spectrum and cause visible avoided crossings~\cite{SupplementalMaterial}. This allows us to qualitatively understand all the main features in the spectra, and to extract quantities of interest. We obtain $E_J/h=14.9(2)$~GHz, $E_L/h=22.2(3)$~GHz, and $E_C/h=0.245(4)$~GHz, confirming that our IST is in the single-well (plasmonic) regime $E_L/E_J>1$ and has a large transmon-like ratio $E_J/E_C\gg1$. A full table of device parameters is reported in the Supplemental Material~\cite{SupplementalMaterial}.

Focusing on the half-flux bias point, we perform standard time-domain characterization of energy relaxation and phase coherence. We obtain mean values of energy relaxation time $T_1=37(3)$~\unit{\micro\second}, Ramsey coherence time $T_2^*=39(4)$~\unit{\micro\second} and echoed coherence time $T_2^E=46(3)$~\unit{\micro\second}, with the standard deviation from repeated measurements quoted in brackets. These are comparable to typical values for tunable transmons in our architecture~\cite{Unpublishedfluxpaper}.

\begin{figure}[t!]
\includegraphics[width=8.5cm]{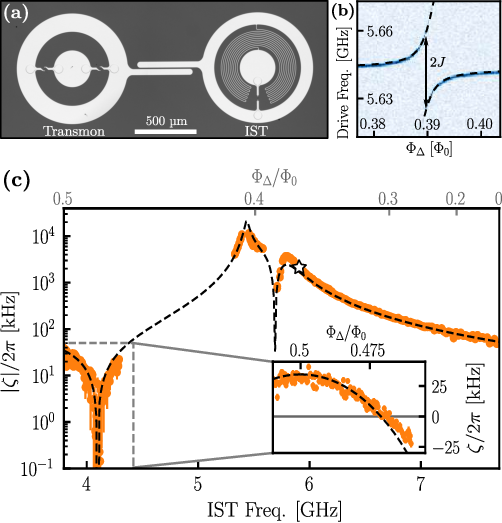} 
\caption{\label{fig:IST_Tr_Device} (a) Optical micrograph of device B, consisting of a tunable gradiometric transmon coupled to an IST via capacitive arms. (b) Spectroscopy of the avoided crossing obtained by sweeping the IST frequency through resonance with the transmon. A fit to the data gives $J/2\pi$ = 17.15(5) MHz (dashed lines). (c) Magnitude of static ZZ shift $\left|\zeta\right|$ across the entire IST frequency range. The x-axis is chosen to be linear in frequency (bottom) instead of flux bias (top). Experimental data shown as orange points, numerical model as black dashed line. Inset shows the region around the zero crossing of $\zeta$ on a linear scale. The star denotes the operation point for sideband gates (see main text).}
\end{figure}

Next, we investigate a system of two qubits with opposite anharmonicity, with the goal of suppressing the unwanted ZZ interaction. We fabricate and characterize device B, consisting of an IST coupled directly to a gradiometric transmon via a static mutual capacitance, as shown in Fig~\ref{fig:IST_Tr_Device}~(a). Qubits are controlled through individual XYZ lines and read out through individual coaxial resonators. Circuit parameters for device B are obtained from spectroscopy as described earlier, and are also found in the Supplemental Material~\cite{SupplementalMaterial}. Although we use a tunable transmon in this particular device, it is kept at integer flux bias throughout this work and therefore at its maximum frequency of 5.64~GHz. The IST has a minimum and maximum frequency of 3.79~GHz and 7.71~GHz, respectively. The capacitive arms lead to a direct coupling strength $J/2\pi=17.15(5)$~MHz at the transmon sweet spot frequency, obtained from the spectroscopy data in Fig~\ref{fig:IST_Tr_Device}~(b).

To quantify the amount of unwanted static ZZ shift arising from the fixed coupling $J$, we perform conditional Ramsey fringe measurements on the transmon with the IST prepared either in its ground or excited state. The shift in transmon frequency directly provides us with the ZZ interaction strength $\zeta$ defined in Eq.~(\ref{eq:zeta_def}). This experiment is repeated as a function of IST frequency across its entire flux range, and the resulting values of $\zeta$ are plotted in Fig.~\ref{fig:IST_Tr_Device}~(c). At $\Phi_\Delta = 0.5\Phi_0$, we find a small but finite ZZ shift of  $+27(2)$~kHz. Varying the IST bias quickly leads to a reduction of $\alpha_{\textrm{IST}}$, which causes $\zeta$ to cross zero and then change sign. The optimal point is found at $\Phi_\Delta \approx 0.47\Phi_0$ in this device, with the ZZ interaction effectively suppressed to zero within the coherence time of the qubits, setting a bound $|\zeta|/2\pi<5$~kHz. Interestingly, we find that here $\alpha_{\textrm{IST}}/2\pi\approx150$~MHz is significantly smaller than the transmon anharmonicity, suggesting that the simple condition $\alpha_{\textrm{IST}} = - \alpha_{\textrm{Tr}}$ is insufficient for predicting the optimal ZZ suppression point in practice. This is because the Duffing oscillator approximation neglects the effect of all higher non-computational levels $\ket{\widetilde{ij}}$ with $i+j>2$, as well as the existence of odd-parity terms in the IST Hamiltonian that appear away from flux sweet spots (as discussed later).

To accurately model the ZZ interaction strength versus flux, we numerically diagonalize the full two-qubit Hamiltonian to extract its energy spectrum and then calculate $\zeta$ directly using its definition in Eq.~(\ref{eq:zeta_def})~\cite{SupplementalMaterial}. All quantities in the Hamiltonian, including $J$, have been independently determined from spectroscopy, leaving no free fitting parameters. The numerical result, overlaid to the data in Fig.~\ref{fig:IST_Tr_Device}~(c), shows excellent agreement with the experiment over the entire flux range and over many orders of magnitude in $|\zeta|$. This indicates that we are capturing the essential physics of this system, and that our method can be used to quantitatively predict the ZZ interaction in future devices with modified parameters. Further simulations show that $\zeta=0$ can be achieved at the flux sweet spot and at arbitrary IST-transmon detuning $\Delta$ simply by adjusting the Josephson energies of the two qubits~\cite{SupplementalMaterial}. This would allow for the straightforward targeting of optimal regimes for conventional two-qubit gate schemes developed for transmons, such as the all-microwave cross-resonance gate~\cite{rigetti2010CR, sheldon2016CR}, fast-flux unipolar or bipolar gates~\cite{barends2019diabatic, negirneac2021SNZ}, and parametric gates~\cite{reagor2018parametric, sete2021parametricresonance}, while minimizing both coherent and incoherent errors.

\begin{figure}[t]
\includegraphics[width=8.5cm]{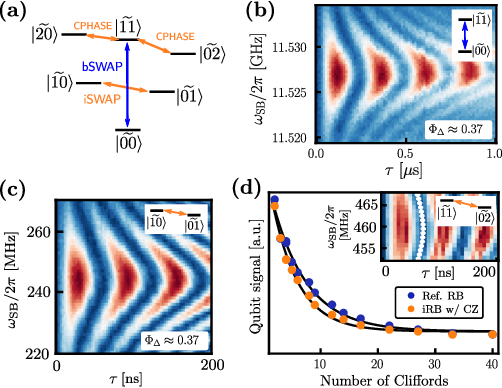} 
\caption{\label{fig:Sidebands} (a) Qualitative energy level diagram of the coupled IST-transmon system, showing the main sideband transitions allowed by the IST cubic non-linearity and the corresponding labels for two-qubit unitary operations. (b)-(c) Coherent Rabi oscillations between the two states indicated in the insets, as a function of frequency $\omega_{\mathrm{SB}}$ and duration $\tau$ of the sideband pulse applied to the IST. Shown is the population of the transmon (ground state in blue, excited state in red). (d) Interleaved randomized benchmarking sequence of a CZ gate based on the red sideband transition $|\widetilde{11}\rangle \leftrightarrow |\widetilde{02}\rangle$. Plotted here is the data obtained from the transmon. Inset shows the Rabi chevron pattern of the transition, with the white dots indicating the optimal gate time for a complete return to the $|\widetilde{11}\rangle$ state.}
\end{figure}

A feature of the RF-SQUID Hamiltonian in Eq.~(\ref{eq:RF-SQUID_Hamiltonian}) is that biasing away from the symmetry points introduces odd-power terms in the expansion of the phase potential $U(\hat{\phi})$. More specifically, cubic phase terms $\sim\hat{\phi}^3$ can lead to useful new interactions, which have been previously used for parametric amplification with three-wave mixing~\cite{Frattini2017} and for two-qubit sideband interactions in a CSFQ-transmon circuit~\cite{noguchi_fast_2020}. We further explore the latter using our IST-transmon system. Fig.~\ref{fig:Sidebands}~(a) shows a qualitative level diagram with the main transitions of interest that can be driven by applying an RF drive to the IST, close to either the difference frequency (red sideband) or sum frequency (blue sideband) of two individual qubit transitions. Note that these can be driven as single-photon transitions via the cubic non-linearity, which is not possible in a transmon-transmon system~\cite{SupplementalMaterial}. 

We demonstrate these transitions by biasing the IST to $\Phi_\Delta \approx 0.37 \Phi_0$, turning on the cubic nonlinearity. Here, the IST frequency is close to 5.9 GHz, slightly above the transmon, and its anharmonicity is $\alpha_{\textrm{IST}}/2\pi\approx -100$~MHz. We apply a sideband tone of frequency $\omega_{SB}$ to the  IST only, with the pulse envelope consisting of Gaussian rise and fall edges of width 15~ns each and a plateau of variable duration $\tau$. By sweeping $\omega_{SB}$ around the frequency of the $|\widetilde{00}\rangle \rightarrow |\widetilde{11}\rangle$ transition, we observe a typical chevron pattern of Rabi oscillations in both qubits simultaneously, as shown in Fig.~\ref{fig:Sidebands}~(b) for the transmon. This blue sideband interaction forms the primitive for a bSWAP unitary~\cite{poletto_entanglement_2012-1, wei_native_2023}. We can fully swap the $|\widetilde{00}\rangle $ and $|\widetilde{11}\rangle$ populations within 125~ns, or create a Bell state $[|\widetilde{00}\rangle + |\widetilde{11}\rangle]/ \sqrt{2}$ by shortening the pulse accordingly. As an example of a red sideband, we can instead prepare the system in $|\widetilde{01}\rangle$ and drive Rabi oscillations to $|\widetilde{10}\rangle$, as shown in Fig.~\ref{fig:Sidebands}~(c). 

We now follow the procedure outlined in Ref.~\cite{noguchi_fast_2020} to calibrate a red sideband gate using the $|\widetilde{11}\rangle \rightarrow |\widetilde{02}\rangle$ transition, implementing a CPHASE$(\pi)=$~CZ unitary. This includes an active cancellation tone to remove the large static ZZ interaction at this bias point (see star in Fig.~\ref{fig:IST_Tr_Device}). Driving the sideband of interest for the correct duration, as shown in the inset of Fig.~\ref{fig:Sidebands}~(d), returns the population to the computational state $|\widetilde{11}\rangle$ but with an acquired relative phase. Local single-qubit phases caused by ac-Stark shifts on other states during the sideband drive are removed in software using virtual Z gates~\cite{McKay2017} on each qubit. The resulting CZ gate has a total gate time $t_g = 75$~ns. We analyze its performance using interleaved randomized benchmarking~\cite{magesan_efficient_2012-1, corcoles_process_2013}, as shown in Fig.~\ref{fig:Sidebands}~(d). Averaging the measured data from both IST and transmon we obtain a gate fidelity of $F_{CZ}=95.8(1.3)$\%, mainly limited by the short coherence time of the IST at this bias point due to flux noise ($T_2^*\sim 0.4$~\unit{\micro\second}, $T_2^E\sim 1.2$~\unit{\micro\second}), similar to the experiment in Ref.~\cite{noguchi_fast_2020}. Separately, we extract the amplitude of 1/f flux noise $A_\phi$ for the IST and find $A_\phi = 6.8(0.1)$~\unit{\micro\Phi_0}, comparable to our transmons~\cite{SupplementalMaterial}. The observed strong reduction in coherence can be mainly explained by the steep frequency-flux curve of the IST, which could be reduced in the future by adjusting circuit parameters to achieve a shallower flux dependence.

In summary, we experimentally realized a transmon shunted by a geometric linear inductor and showed suppressed ZZ interaction when coupling it to a conventional transmon, as well as the existence of several two-qubit sideband interactions. Its ease of fabrication and integration into an existing transmon-based platform make the IST a promising complementary qubit species. Its single-JJ nature also makes it compatible with post-fabrication annealing techniques~\cite{IBMannealingJJ, OQCannealingJJ}, ensuring reproducibility at larger scales. The IST could be employed in multi-level qudit schemes thanks to its built-in protection from charge noise~\cite{koch_charging_2009, Peterer2015}, or in the study of strongly driven systems due to the steep parabolic confinement provided by $E_L>E_J$, preventing chaotic dynamics~\cite{verney2019structural, chaotictransmon2023}.


\begin{acknowledgments}

 This work has received funding from the United Kingdom Engineering and Physical Sciences Research Council (EPSRC) under Grants No. EP/J013501/1, No. EP/N015118/1, and No. EP/T001062/1. SDF acknowledges support from the Swiss Study Foundation. MB acknowledges support from an EPSRC Quantum Technology Fellowship, Grant No. EP/W027992/1. SC acknowledges support from an Eric and Wendy Schmidt AI in Science Postdoctoral Fellowship. The authors also acknowledge related earlier work by Y.~Nakamura~\textit{et al.}, circulated at the American Physical Society (APS) March Meeting 2021.
 
\end{acknowledgments}

\providecommand{\noopsort}[1]{}\providecommand{\singleletter}[1]{#1}%

\end{document}



\title{Complementing the transmon by integrating a geometric \\ shunt inductor \\-- Supplemental Material --}

\author{Simone D. Fasciati}
\email[Corresponding author: ]{simone.fasciati@physics.ox.ac.uk}
\author{Boris Shteynas}
\author{Giulio Campanaro}
\author{Mustafa Bakr}
\author{Shuxiang Cao}
\author{Vivek Chidambaram}
\author{James Wills}
\author{Peter J. Leek}
\email[Corresponding author: ]{peter.leek@physics.ox.ac.uk}
\affiliation{Department of Physics, Clarendon Laboratory, University of Oxford, Oxford OX1 3PU, UK}

\date{\today}

\maketitle

\tableofcontents
\newpage

\section{\label{sec:app_theory} Theory and derivations}

\subsection{ZZ interaction in the Duffing oscillator picture}

Two weakly-anharmonic Duffing oscillators experience a cross-Kerr (ZZ) shift $\zeta$ when coupled to each other. The ZZ shift shows up as the term of the Hamiltonian diagonal in the number basis. Here we briefly derive the standard perturbative expression for $\zeta$.

The Hamiltonian of two coupled Duffing oscillators, characterized by the frequencies $\omega_1$ and $\omega_2$, anharmonicities $\alpha_1$ and $\alpha_2$, and coupling strength $J$, written in terms of the creation (annihilation) operators $\hat{a}^{\dagger}$~($\hat{a}$) and $\hat{b}^{\dagger}$~($\hat{b}$) as

$$\hat{H}_0=\omega_1 \hat{a}^{\dagger}\hat{a}+\frac{\alpha_1}{2}\hat{a}^{\dagger}\hat{a}^{\dagger}\hat{a}\hat{a}+\omega_2\hat{b}^{\dagger}\hat{b}+
\frac{\alpha_2}{2}\hat{b}^{\dagger}\hat{b}^{\dagger}\hat{b}\hat{b} + J(\hat{a}^{\dagger}\hat{b}+\hat{a}\hat{b}^{\dagger}),
$$
can be perturbatively diagonalized by a Schrieffer-Wolf transformation $\hat{\tilde{H}} = e^{\hat{S}} \hat{H} e^{-\hat{S}}$, generated by $\hat{S} = p\hat{a}^{\dagger}\hat{b}-p^*\hat{a}\hat{b}^{\dagger}$.
To lowest order, the part which is diagonal in the number basis of the Hamiltonian is

\begin{equation*}
\hat{H}_{diag}\approx\tilde{\omega}_1 \hat{a}^{\dagger}\hat{a}+\frac{\tilde{\alpha}_1}{2}\hat{a}^{\dagger}\hat{a}^{\dagger}\hat{a}\hat{a}+\tilde{\omega}_2\hat{b}^{\dagger}\hat{b}
+\frac{\tilde{\alpha}_2}{2}\hat{b}^{\dagger}\hat{b}^{\dagger}\hat{b}\hat{b} + \frac{2J^2}{\Delta^2}(\alpha_1+\alpha_2)\hat{a}^{\dagger}\hat{a}\hat{b}^{\dagger}\hat{b},
\end{equation*}
where the last term is the cross-Kerr interaction with strength $\zeta=2J^2(\alpha_1 + \alpha_2)/\Delta^2$, and with dressed frequencies $\tilde{\omega}_1$, $\tilde{\omega}_2$ and anharmonicities $\tilde{\alpha}_1$, $\tilde{\alpha}_2$.

\subsection{Perturbative treatment of the IST Hamiltonian at flux sweet spots}

The inductively shunted transmon (IST) presented in this work can be described by the RF-SQUID Hamiltonian (Eq.~2 in the main text)

$$\hat{H} = 4E_C\hat{n}^2-E_J\cos(\hat{\phi}+\phi_e)+\frac{1}{2}E_L\hat{\phi}^2,$$
where $E_J$ is the Josephson energy, $E_L$ is the inductive energy of the linear geometric inductor and $E_C$ is the charging energy of the shunting capacitance. An applied external magnetic flux $\Phi$ is represented here by the reduced flux variable $\phi_e=2\pi\Phi/\Phi_0$. $E_J$ and $E_L$ can be expressed through the equivalent inductances of the junction and the linear inductor as $E_J=\frac{\Phi_0^2}{4\pi^2L_J}$, $E_L=\frac{\Phi_0^2}{4\pi^2L}$. We are interested in the regime of a weak inductive shunt with $E_L>E_J$, or equivalently $L<L_J$, leading to a strong harmonic confinement and therefore to a single-well phase potential (as opposed to the typical corrugated phase potential of the $E_L<<E_J$ limit).  In the vicinity of $\phi_e = \pi$, which corresponds to biasing the RF-SQUID loop with a half flux quantum, one can expand the Hamiltonian as
$$\hat{H}\approx 4E_C \hat{n}^2+\frac{1}{2}(E_L-E_J)\hat{\phi}^2+\frac{1}{24}E_J\hat{\phi}^4.$$

The first two terms of the Hamiltonian form a harmonic oscillator, with the magnitude of the zero-point fluctuations equal to

$$\phi_{zpf} = (2E_C/(E_L-E_J))^{1/4}.$$

The quartic part of the Hamiltonian can be treated as a perturbation, giving a first-order correction 

$$\frac{1}{24}E_J\phi_{zpf}^4\bra{n}(a+a^{\dagger})^4\ket{n}=\frac{1}{24}E_J\phi_{zpf}^4(6n^2+6n+3)$$

to the eigenenergies of the harmonic oscillator Fock states $\ket{n}$.

Thus, the first transition (qubit) frequency and anharmonicity can be written as:

$$\omega_{01}\approx\sqrt{8E_C(E_L-E_J)}+E_C\frac{E_J}{E_L-E_J},$$
$$\alpha \approx\frac{E_J}{E_L-E_J}E_C>0. $$

We therefore obtain a weakly anharmonic circuit with positive values of $\alpha$, matching the experimental results of the main text. One can enhance the anharmonicity beyond typical transmon values $~\alpha~\sim E_C$ by letting $E_L \rightarrow E_J$ and achieving a purely fourth-order (quartic) potential. The perturbative formulas break down in this \textit{quarton} limit, but it is possible to calculate that one indeed obtains a highly anharmonic single-well potential with $\alpha/\omega_{01}\approx33\%$~\cite{Yan2020}.

One can compare these expressions to the other sweet spot at $\phi_e = 0$, where the quadratic terms from the harmonic and cosinusoidal potentials add up "in phase" instead of being subtracted from each other. An analogous derivation as before results in 

$$\omega_{01}\approx\sqrt{8E_C(E_L+E_J)}+E_C\frac{E_J}{E_L+E_J},$$
$$\alpha \approx-\frac{E_J}{E_L+E_J}E_C .$$

Here the IST is again a weakly-anharmonic oscillator, but now with negative anharmonicity like a standard transmon. Note that $|\alpha|< E_C$, resulting in an even weaker anharmonicity than for a transmon with comparable $E_J$ and $E_C$.

\subsection{IST-transmon dynamics and side-band transitions}

The Hamiltonian of the IST with asymmetric potential when detuned from the flux sweet spot and coupled to the standard transmon is (see also Ref.~\cite{noguchi_fast_2020})
$$\hat{H} = \omega_a \hat{a}^{\dagger}\hat{a} + c_3(\hat{a}^{\dagger}+\hat{a})^3 + \frac{\alpha}{2}\hat{a}^{\dagger}\hat{a}^{\dagger}\hat{a}\hat{a} + J(\hat{a}^{\dagger}\hat{b}+\hat{a}\hat{b}^{\dagger})+ \omega_b \hat{b}^{\dagger}\hat{b} + \frac{\beta}{2}\hat{b}^{\dagger}\hat{b}^{\dagger}\hat{b}\hat{b}.$$
The creation (annihilation) operators $\hat{a}^{\dagger} (\hat{a})$ and $\hat{b}^{\dagger} (\hat{b})$ correspond to the IST and the transmon, respectively, and are obtained from the harmonic confinement around the potential minima of their potentials. $\omega_a$, $\omega_b$ and $\alpha$, $\beta$ are the frequencies and anharmonicities of the two nonlinear oscillator. The strength of the cubic nonlinearity of the IST potential is characterized by $c_3$. The two qubits are coupled with exchange interaction with the strength $J$.

The interplay of the cubic nonlinearity $c_3$ and the exchange coupling $J$ produces the interaction between the qubits that can be employed to drive single-photon side-band transitions between two modes. To identify these interaction we need to partially diagonalize the Hamiltonian and eliminate the terms linear with $c_3$ and $J$.

We will search for a Schrieffer-Wolff transformation generated by $\hat{S}$ defined as
$$\hat{S} = d(\hat{a}-\hat{a}^{\dagger}) + p(\hat{a}^{\dagger}\hat{a}^{\dagger}\hat{a}-\hat{a}^{\dagger}\hat{a}\hat{a}) + q(\hat{a}^{\dagger}\hat{b}-\hat{a}\hat{b}^{\dagger}),$$
where the parameters $d$, $p$ and $q$ are chosen such that the linear part (with respect to $J$ and $c_3$) of the Hamiltonian is cancelled. The suggested transformation
$$e^{\hat{S}}\hat{H}e^{-\hat{S}} = \hat{H}+[\hat{S},\hat{H}]+\frac{1}{2}[\hat{S}[\hat{S},\hat{H}]]...$$
will also produce various higher order terms. Since we are only interested in the terms proportional to $c_3 J$, all other terms of higher perturbative orders in $c_3$ and $J$ and higher power of interactions in $\hat{a}^{\dagger}(\hat{a})$ and $\hat{b}^{\dagger}(\hat{b})$ will be omitted. 

To simplify the calculation we find how $\hat{a}$ and $\hat{b}$ are modified by the transformation $\hat{S}$
$$\hat{a} \rightarrow \hat{a}+d+p\hat{a}^2-2p \hat{a}^{\dagger}\hat{a} - q\hat{b} + qp(\hat{a}^{\dagger}\hat{b}+\hat{a}\hat{b}^{\dagger}-\hat{a}\hat{b}) ...$$
$$\hat{b} \rightarrow \hat{b} +q\hat{a} + \frac{1}{2}dq - pq \hat{a}^{\dagger}\hat{a} + \frac{1}{2}pq\hat{a}^2 ...$$

The parameters $d, p, q$ are defined such that they cancel the interactions proportional to $\hat{a}^{\dagger} + \hat{a}$, $\hat{a}^{\dagger}\hat{a}^{\dagger}\hat{a}+\hat{a}^{\dagger}\hat{a}\hat{a}$, and $ \hat{a}^{\dagger}\hat{b}+\hat{a}\hat{b}^{\dagger}$. Under this condition with $\Delta = \omega_a - \omega_b$

$$d = - \frac{3c_3}{\omega_a}, \;\; p = -\frac{2c_3 (1-\frac{3}{2}\frac{\alpha}{\omega_a})}{\omega_a - \alpha}, \;\; q = \frac{J}{\Delta}.$$

After we cancel the linear order terms in the original Hamiltonian and omit all the higher order terms, the transformed Hamiltonian includes two new interactions with the strength $M_1$ and $M_2$

\begin{equation*}
\hat{\tilde{H}} = \tilde{\omega}_a \hat{a}
^{\dagger}\hat{a} + \frac{\alpha}{2}\hat{a}^{\dagger}\hat{a}^{\dagger}\hat{a}\hat{a} + \tilde{\omega}_b \hat{b}^{\dagger}\hat{b} + \frac{\beta}{2}\hat{b}^{\dagger}\hat{b}^{\dagger}\hat{b}\hat{b} +
M_1 \hat{a}^{\dagger}\hat{a} (\hat{b}+\hat{b}^{\dagger}) + M_2 (\hat{a}^{\dagger 2}\hat{b} + \hat{a}^2 \hat{b}^{\dagger}),
\end{equation*}
where $M_1 = (2\omega_a-\omega_b)pq - 4c_3q - 2Jp - 2\alpha dq $ and $ M_2 = \frac{1}{2}\omega_b p q - 5c_3q + Jp - \alpha dq$. The new renormalized frequencies $\tilde{\omega}_a$ and $\tilde{\omega}_b$ differ from the original oscaillator frequencies $\omega_a$ and $\omega_b$, and correspond to the observable qubit frequencies measured experimentally.

After the substitution of the obtained expressions for $d, p$ and $q$ we arrive at

$$M_1 = \frac{2c_3J}{\Delta}\Big(\frac{-3\omega_a+\Delta+2\alpha}{\omega_a - \alpha} + \frac{3\alpha}{\omega_a}(1+\frac{1}{2}\frac{\omega_a-\Delta}{\omega_a-\alpha})\Big)$$

$$M_2 = \frac{c_3J}{\Delta}\Big(\frac{-6\omega_a-\Delta+5\alpha}{\omega_a-\alpha} + \frac{3\alpha}{\omega_a}(1+\frac{1}{2}\frac{\omega_a+\Delta}{\omega_a-\alpha})\Big)$$
In the limiting case when $\omega_a \gg \Delta, \alpha$ the strength of the interactions scales as
$$M_1, M_2 \sim - 6 c_3J/\Delta$$

The terms $M_1 a^{\dagger}a (b+b^{\dagger})$ and $ M_2 (a^{\dagger 2}b + a^2 b^{\dagger})$, since they are quadratic in $a^{\dagger}(a)$ and linear in $b^{\dagger}(b)$, can be used for generation of the side-band transitions between the modes $a$ and $b$. With the IST drive $H_d = \Omega(a^{\dagger}e^{-i\omega_d t} + a e^{i\omega_d t})$, where $\Omega$ is the effective Rabi frequency and $\omega_d$ is the drive frequency, the strength of the resonant side-band interaction is going to be proportional to $\frac{\Omega}{\omega_d-\tilde{\omega}_a}(M_1 + M_2)$.

\section{\label{sec:app_fab_parameters} Fabrication and device parameters}

We fabricate devices on 3-inch, double-side polished, intrinsic (resistivity $\rho>10$~k$\Omega$) 100-orientation silicon wafers. The native oxide is first removed with a buffered oxide etch, and the substrate is then immediately transferred to vacuum inside a Plassys MEB550~S2 electron-beam evaporation system. We deposit 100 nm of aluminum on one side of the wafer, after which we unload it, flip it, and load it again to perform the same deposition on the opposite side. We then carry out a single step of standard photolithography on each side of the substrate to define the coaxial qubit electrodes and readout resonators, respectively. One side is always covered in a protective layer of photoresist while the other is being processed. Wet etching of the previously deposited aluminum film is used to define all large circuit features, including the IST meandering inductors. Subsequently, Josephson junctions are patterned using electron-beam lithography and deposited using a conventional Dolan-bridge method and double-angle evaporation of aluminum. Large overlap areas between the junction leads and the previously defined features, along with in-situ ion milling before metal deposition, ensure a low contact resistance. Finally, the wafer is diced into $5\times 5$~mm$^2$ dies, which are then cleaned and selected for cryogenic characterization.

The main experimentally obtained parameters for devices A and B presented in this work are summarized in Table~\ref{tab:device_overview}. Most quantities are valid for qubits biased at their respective sweet spots, i.e. the transmon is biased at integer flux quantum (maximum frequency), while the ISTs are biased at half flux quantum (minimum frequency). The main exception is the exchange coupling $J$, which is measured in spectroscopy by sweeping one qubit through the sweet spot frequency of the other, and observing an avoided crossing. The mutual inductance $M = \Phi/I$ between a qubit and its respective XYZ control line is obtained from the applied DC current $I$ required to observe one full period of oscillation (equivalent to a flux quantum $\Phi_0$) in qubit and resonator spectroscopy.

To extract all Hamiltonian circuit parameters, flux-dependent spectroscopy across the entire flux range is carried out and the resulting data is fitted to numerical spectra of the single-qubit Hamiltonians, including the readout resonator mode as well as parasitic inductor modes for the IST. The qubits on device B are treated individually and the coupling $J$ is ignored in this step, as it does not appreciably affect the transmon and IST spectra. See Supplemental section~\ref{sec:app_spectroscopy_fit} for more details on the fitting method. The transmon used here is a tunable DC-SQUID transmon consisting of two JJs with energies $E_{J,1}$ and $E_{J,2}$, which are reported here in terms of the total Josephson energy $E_J = E_{J,1} + E_{J,2}$ and the asymmetry $d=|E_{J,1} - E_{J,2}|/E_{J}$.

We also include the number of repetitions and the total run time for measurements of relaxation times $T_1$ and coherence times $T_2^*$ and $T_{2,E}$. The spread $\sigma_f$ of all frequency values measured from the repeated Ramsey traces during this time is also reported, to show typical frequency fluctuations over longer timescales.

\begin{table*}[t!]
    \begin{spacing}{1.2}
    \centering
    \caption{\label{tab:device_overview} Summary of parameters for devices A and B presented in this work. Basic quantities are obtained when qubits are biased at their respective flux sweet spots ($\phi_{e}=0$ for the transmon, $\phi_{e}=0.5$ for the IST) except for the exchange couplings $J$ in device B. Numerical fits to spectroscopy data as a function of flux are performed to extract circuit parameters. Numbers in brackets denote measurement uncertainty for measured quantities, fit uncertainty for numerically extracted parameters, and statistical variation for the repeated coherence measurements.}

    \begin{tabular}{ccrccw{c}{4cm}w{c}{2cm}w{c}{2cm}} \toprule
            & & & & &  Device A &  \multicolumn{2}{c}{Device B}  \\ \midrule
           & & Quantity & Unit & &  IST &  Transmon & IST  \\ \midrule
          \multirow{6}{*}{\rotatebox[origin=c]{90}{\textit{Basic quantities}}} &  & Qubit frequency $\omega_{01}/2\pi$ &(GHz)& & 3.659 & 5.644 & 3.785 \\
          & & Anharmonicity $\alpha/2\pi$ &(MHz) &   & $+$228 & $-$220 & $+$281\\
          & & Resonator frequency $\omega_r/2\pi$ &(GHz) & & 9.730 & 9.843 & 9.969 \\
          & & Resonator linewidth $\kappa/2\pi$ &(MHz) &  & 0.76(6) & 0.52(5) & 0.53(1) \\
          & & Dispersive shift $2\chi/2\pi$ &(MHz) &  & +0.9(1) & $-$1.5(1) & +1.2(1) \\ 
          & & Mutual inductance $M$ &(pH) &  & 0.97 & 1.71 & 1.10 \\\midrule
          \multirow{7}{*}{\rotatebox[origin=c]{90}{\textit{Numerical fit}}} & & Charging energy $E_C/h$ &(GHz) & & 0.245(4) & 0.203(2) & 0.238(1) \\
           & & Total Josephson energy $E_J/h$ &(GHz) & & 14.9(2) & 21.2(2) & 19.4(1) \\
           & & Inductive energy $E_L/h$ &(GHz) & & 22.2(3) & -- & 27.2(1)\\
           & & DC SQUID asymmetry $d$ & & & -- & 0.20(2) & -- \\
           & & Ratio $E_J/E_C$ & & & 61 & 104 & 82\\ 
           & & Inductance ratio $E_L/E_J$ & & & 1.5 & -- & 1.4 \\ 
           & & Resonator coupling $g_r/2\pi$ &(MHz) & & 180 (10) & 154(1) & 201(10) \\\midrule
          \multirow{6}{*}{\rotatebox[origin=c]{90}{\textit{Coherence stats.}}} & & Relaxation time $T_1$ &(\unit{\micro\second}) &   & 37(3) &  38(9)  & 47(5)  \\
          & & Ramsey coherence time $T_2^*$ &(\unit{\micro\second}) &  &  39(4) &  18(2)  & 12(2)  \\
          & & Hahn-echo coherence time $T_{2,E}$ &(\unit{\micro\second}) &  & 46(3) &  26(3)  & 23(2)  \\
          & & Frequency variation $\sigma_f$ &(kHz) &  & 0.25 & 5.5 & 7.2  \\
          & & \# of repetitions & &  & 10 & 130 & 96 \\
          & & Total run time &(h) & & 0.5 & 11.2 & 8.8 \\ \midrule
          \multirow{4}{*}{\rotatebox[origin=c]{90}{\textit{Coupling}}} & & \multirow{2}{*}{Exchange coupling $J/2\pi$ } & \multirow{2}{*}{(MHz)}& & -- & \multicolumn{2}{c}{10.93(4) at $\sim$ 3.79 GHz} \\
            & & & & & -- & \multicolumn{2}{c}{17.15(5) at $\sim$ 5.64 GHz} \\
            & & Fixed cap. coupling $g_c/2\pi$ &(MHz) & & -- & \multicolumn{2}{c}{ 9.7(1) } \\
            & & Static ZZ shift $\zeta/2\pi$ &(kHz) & & -- & \multicolumn{2}{c}{27(2)} \\

         \bottomrule
    \end{tabular}
    \end{spacing}
    
\end{table*}

\section{\label{sec:app_setup}Experimental setup}

\begin{figure*}[t!]
\includegraphics[width=16cm]{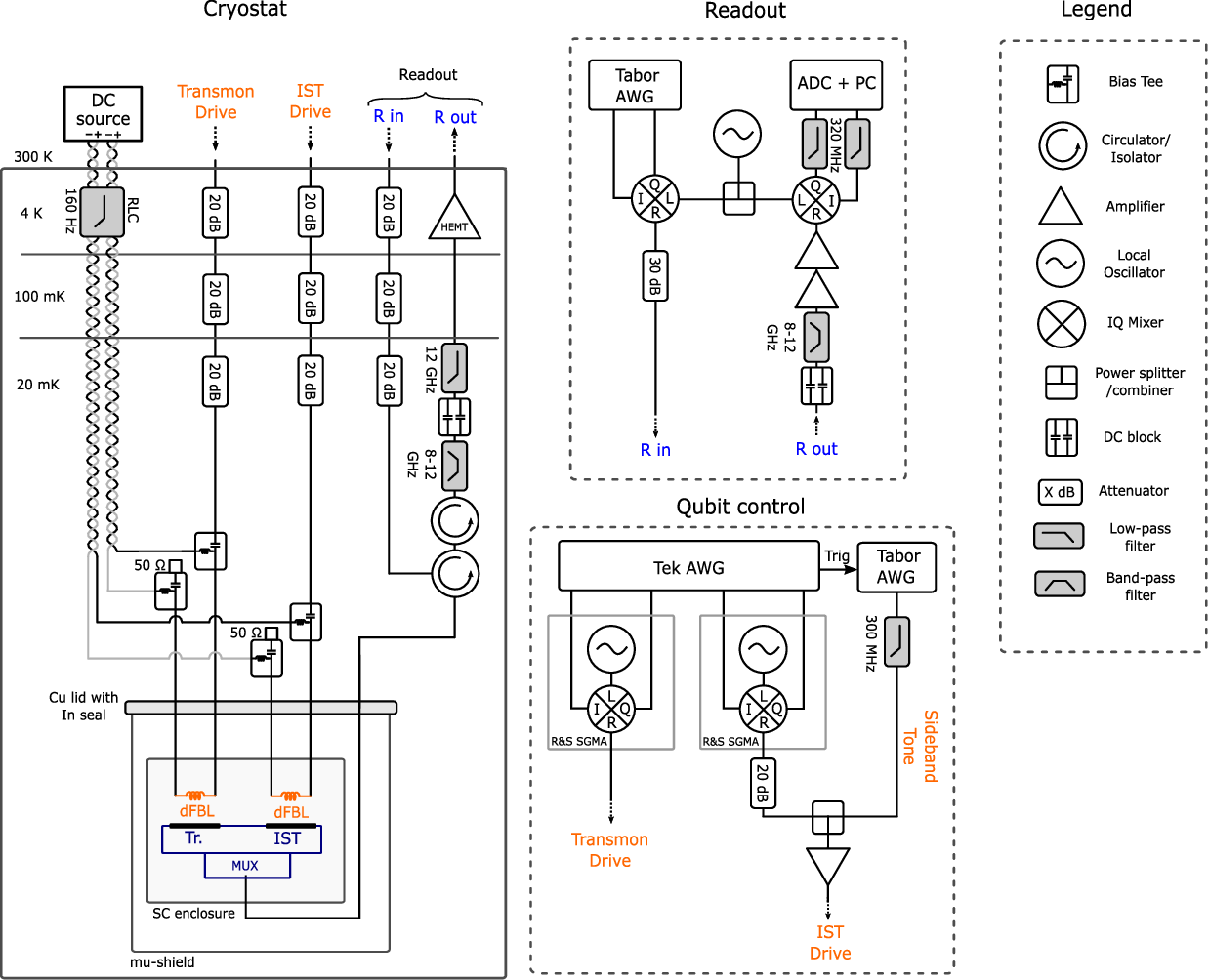} 
\caption{\label{fig:cryosetup} Schematic drawing of the cryogenic measurement setup used in this work. The configuration shown was used for the measurement of device B (coupled transmon-IST system). For device A (individual IST), measured in a separate cooldown, the wiring is analogous but with only a single qubit drive line used and without the additional AWG channel for sideband drives.}
\end{figure*}

The experiments in this work are carried out using the cryogenic measurement setup shown schematically in Fig.~\ref{fig:cryosetup}. The sample enclosure housing the device to be measured is installed at the base plate of a dilution cryostat, with individual control lines for each qubit and a common readout chain for multiplexed readout. We utilize a 3D-integrated readout multiplexing cavity that can couple to up to four resonators simultaneously~\cite{UnpublishedMUXpaper}. Qubit microwave drives and DC flux control are filtered and attenuated separately, and then combined at base temperature using bias tees. Signals reach the qubits via off-chip differential flux bias lines (dFBL), which are built into the sample enclosure and can provide full XYZ control of gradiometric qubits with low crosstalk~\cite{Unpublishedfluxpaper}. For the transmon-IST sideband experiments, an additional arbitrary waveform generator (AWG) channel is used to create an RF tone up to several 100 MHz, which is then combined with the standard IST microwave drive at room temperature using a wideband (resistive) power combiner. This allows us to drive the sideband transitions while also still performing single qubit gates on the IST. Due to the voltage output limit of the AWG and the total amount of attenuation on the drive line, we had to amplify the signal at room temperature after the power combiner in order to drive the sideband transitions strongly enough. This limitation could be overcome in the future by adjusting the attenuation and filtering of the drive line in the cryostat.

\section{\label{sec:app_spectroscopy_fit}Numerical fitting of the IST spectrum}

This section describes in more detail the fitting method used to extract all IST circuit parameters from its flux-dependent frequency spectrum, such as the one shown in Fig.~2 in the main text. This approach is analogous to other realizations of RF-SQUID circuits such as the fluxonium~\cite{manucharyan_fluxonium_2009} or the geometric superinductance qubit~\cite{peruzzo_geometric_2021}. The large distributed nature of the shunting inductor gives rise to unwanted (parasitic) modes in the circuit that can couple strongly to our desired fundamental (lumped) mode and significantly distort its spectrum. This is taken into account by adding these modes to the fitting Hamiltonian 
\begin{equation}
\label{eq:H_parmodes}
    \mathcal{H}_{RF, par} = \mathcal{H}_{RF} + 
    \hbar \omega_{r}\hat{c}_r^\dagger\hat{c}_r + \hbar g_{r}\hat{n}\left(\hat{c}_r^\dagger + \hat{c}_r  \right) + \sum_{j=1}^n \left[ \hbar \omega_{p,j}\hat{c}_j^\dagger\hat{c}_j + \hbar g_{p,j}\hat{n}\left(\hat{c}_j^\dagger + \hat{c}_j  \right) \right], \tag{S1}
\end{equation}
where $\mathcal{H}_{RF}$ is the standard RF-SQUID Hamiltonian from Eq.~(2) in the main text, and  $\hat{c}_j^\dagger$ $ (\hat{c}_j)$ is the creation (annihilation) operator for each linear mode $j$ with frequency $\omega_{p,j}$ and coupling strength $g_{p,j}$ to the IST. We also include the (intentional) coupling to the readout mode described by $\hat{c}_r^\dagger$, $\omega_{r}$ and $g_{r}$. We neglect the coupling $J$ to the transmon mode in the case of device B, as this is at least one order of magnitude smaller than all other coupling strengths in this system. The number $n$ of modes to be included is arbitrary and depends on the details of the system. In practice we find satisfactory results by including only the two lowest modes in the fit, even if in principle there are an infinite number of modes. The spectrum of $\mathcal{H}_{RF,par}$ is then obtained by numerical diagonalization using the QuTiP package~\cite{JOHANSSON20131234}, and fitted to experimental spectroscopy data to obtain all Hamiltonian parameters.

\begin{figure*}[t!]
\includegraphics[width=16cm]{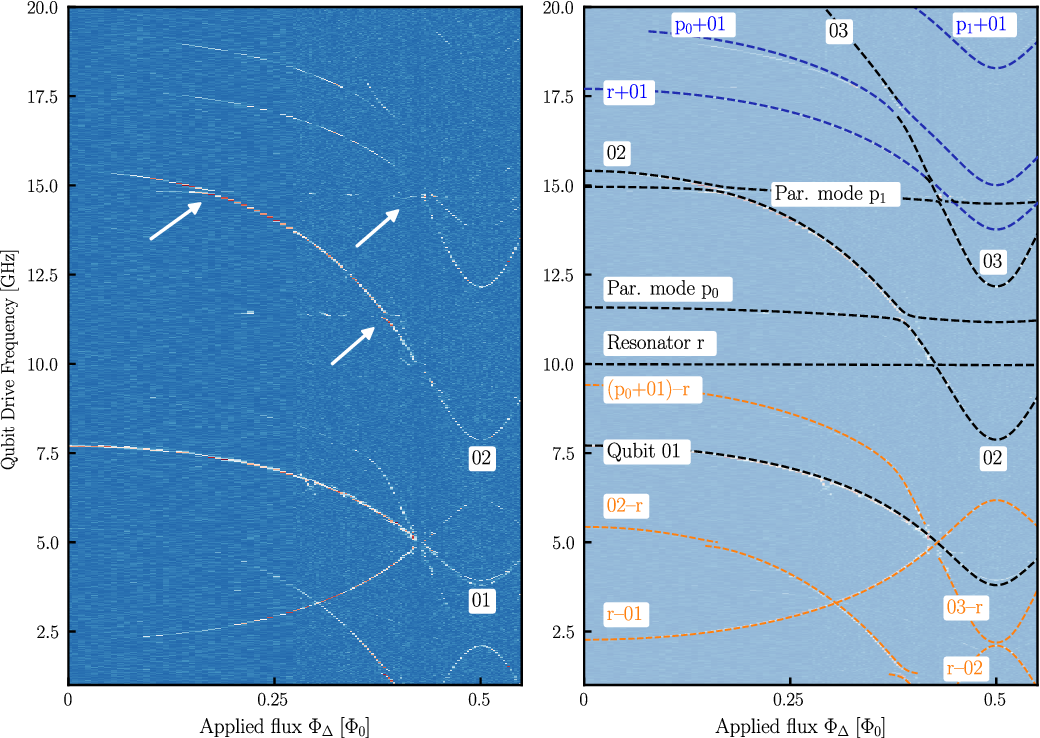} 
\caption{\label{fig:IST_deviceB_spectroscopy} (a) Wide-range two-tone VNA spectroscopy of IST on device B as a function of applied external flux. The two lowest transitions of the IST are labeled as 01 and 02, respectively. Large avoided crossings with additional modes in the system are indicated by white arrows. (b) Same data as in (a), overlaid with a labeled energy spectrum (dashed lines) fitted numerically using the Hamiltonian in Eq.~(\ref{eq:H_parmodes}). We also include sum and difference frequencies (blue and red sidebands, respectively) of several main transitions.}
\end{figure*}

The effect of parasitic modes can be clearly seen when performing wide-frequency continuous-wave spectroscopy, such as for IST B shown in Fig.~\ref{fig:IST_deviceB_spectroscopy}~(a) up to 20 GHz. The first and second transitions of the IST mode can be quickly identified based on approximate knowledge of device parameters, and are here labeled as 01 and 02, respectively. There are many additional features visible due to the relatively high probe power used here, which excites higher-order transitions. Large avoided crossings between the 02 transition and quasi-horizontal modes are visible at frequencies around 11.4 and 14.8 GHz, which we attribute to geometric inductor modes. This can be confirmed by simulating the geometry of the IST circuit in a classical finite-element solver (Ansys HFSS), which predicts the two lowest eigenmodes of the inductor to lie at 11.8 and 15.2~GHz, close to the values observed in the experiment. We therefore include them in the numerical fitting, and extract coupling strengths $g_{p,1}/2\pi\approx 0.9$~GHz and $g_{p,2}/2\pi\approx 1.3$~GHz, along with all remaining device parameters which are reported in Table~\ref{tab:device_overview}. We then verify our fit result by overlaying the numerically obtained spectra to the experimental data, as shown in Fig.~\ref{fig:IST_deviceB_spectroscopy}~(b). The main transitions of our model Hamiltonian are shown as black dashed lines, and while they match the data very well, there are many additional features that remain unlabeled. These can be identified as blue or red sideband transitions between any two (and in one case, three) modes, by simply calculating the sums or differences of the main transitions. We find that most of the possible combination of two frequencies corresponds to an observable feature in the data, and we label them correspondingly (blue and orange dashed lines, respectively). This allows us to fully map out this rich energy spectrum, and suggests that we are capturing all the important physics of the circuit. Also, note that we did not explicitly include the 03 transition frequency in the experimental dataset used as input for the fitting routine for IST device B, but the calculated numerical spectrum indeed matches the experimental 03 transition very well, providing a useful self-consistency check of the model.

Knowledge of sideband transitions enables us to identify the additional feature seen in the high-power spectroscopy of IST device A, labeled "SB", see inset of Fig.~2~(b) in the main text. This corresponds to red sideband (p$_1+01$)$-$r, which is very close to the 01 transition in this device due to the larger designed shunting inductance compared to IST device B and the resulting lower frequency of the first parasitic mode, $\omega_{p,1}/2\pi\approx 10.3$~GHz. 

The extremely large couplings $g_{p,j}/2\pi\sim 1$~GHz to parasitic modes are due to the large field overlap of the coaxial IST mode excitation and the geometric inductor. In the future, reducing the footprint of the inductor and tweaking its design should at the same time reduce $g_{p,j}$ and increase the mode frequencies $\omega_{p,j}$, leading to weaker distortions and fewer collisions with the main IST transitions. This is especially important for the 02 transition, which determines the anharmonicity of the IST (as shown in  Fig~2~(c) in the main text) and is also important in many qubit-qubit interaction schemes, such as the sidebands and the CZ gate studied in this work. In conclusion, we note that the good agreement between model and experiment shown here, as well as the ability to predict the frequency of unwanted geometric modes from a relatively inexpensive classical finite-element simulation, will help inform future design iterations of the IST.

\section{\label{sec:app_ZZ_simulations} Simulation of ZZ interaction in the IST-transmon system}

\subsection{Simulation method and comparison to experiment}

To explore the general behavior of the ZZ interaction in the studied transmon-IST system and to verify our experimental result for the interaction strength $\zeta$, we perform numerical simulations as a function of various circuit parameters. This allows us to obtain the numerical prediction shown in Fig.~3~(c) of the main text. We start from the general Hamiltonian describing an IST and a transmon coupled via a static exchange interaction,

\begin{equation}
    \label{eq:full_2QHam_ZZsim}
    \mathcal{H}_{2Q} = \mathcal{H}_{RF,par} + \mathcal{H}_{CPB} + \mathcal{H}_{int}, \tag{S2}
\end{equation}

with $\mathcal{H}_{RF,par}$ defined in Eq.~(\ref{eq:H_parmodes}), $\mathcal{H}_{CPB}$ the standard Cooper-pair-box (CPB) Hamiltonian for the transmon
\begin{equation}
    \label{eq:H_CPB}
    \mathcal{H}_{CPB} = 4E_{C,Tr}\hat{n}_{Tr}^2-E_{J,Tr}\cos(\hat{\phi}_{Tr}), \tag{S3}
\end{equation}

and the interaction term

\begin{equation}
    \label{eq:H_int}
    \mathcal{H}_{int} = \hbar g_c \hat{n}_{IST} \hat{n}_{Tr} = \hbar J (\hat{a}^\dagger - \hat{a})(\hat{b}^\dagger - \hat{b}). \tag{S4}
\end{equation}

Here, $g_c$ is the bare capacitive coupling strength between the two qubits, $J$ is the (frequency-dependent) effective exchange interaction strength, and we expressed the charge number operators as bosonic ladder operators via the standard relation $\hat{n} = i/(\sqrt{2}\phi_{zpf})\times(\hat{a}^\dagger - \hat{a})$. The zero-point phase fluctuations are defined as $\phi_{zpf,IST}=[8E_{C,IST}/(E_{L,IST}-E_{J,IST})]^{1/4}$ and $\phi_{zpf,Tr}=[8E_{C,Tr}/E_{J,Tr}]^{1/4}$ for the IST and transmon, respectively.

We numerically diagonalize the full Hamiltonian $\mathcal{H}_{2Q}$ in QuTiP~\cite{JOHANSSON20131234} without any approximations to the cosine phase potentials, and with all device parameters fixed to the experimentally determined values for device B, reported in Table~\ref{tab:device_overview}. The only simplification is the exclusion of the readout resonator modes, since they do not appreciably affect the ZZ landscape and their omission significantly shortens the simulation run time. The parasitic inductor modes of the IST are instead included due to their strong coupling, to ensure that the simulated IST spectrum matches the experiment (see also related discussion in Supplemental section~\ref{sec:app_spectroscopy_fit}). The obtained energy levels are then labeled as the dressed two-qubit eigenstates $\ket{\widetilde{ij}}$, and the ZZ interaction strength is extracted as $\zeta = E_{\ket{\widetilde{11}}}-E_{\ket{\widetilde{10}}} - E_{{\ket{\widetilde{01}}}}+E_{\ket{\widetilde{00}}}$ (Eq.~(1) in the main text). This procedure is repeated as a function of the external flux parameter $\phi_e$ in the IST Hamiltonian, in order to obtain the flux dependence $\zeta(\Phi_\Delta)$ as plotted in Fig.~3~(c) in the main text. Note that the labeling $\ket{\widetilde{ij}}$ is ambiguous in the vicinity of avoided crossings in the spectrum, where states are strongly hybridized, but this definition of $\zeta$ is still useful to calculate the interaction strength across the entire range~\cite{zhao_high-contrast_2020}.

\subsection{Further exploration of ZZ landscape}

Given the very close agreement between experiment and numerical prediction, we now want to explore the dependence of $\zeta$ on device parameters that are fixed in the present device B (such as the capacitive, Josephson or inductive energies of the qubits) but which could be modified in a future design iteration. Here we choose, as a representative example, to vary the two quantities $E_{C,IST}$ and $E_{J,IST}$, while keeping all other parameters in the Hamiltonian fixed as before. While this choice seems arbitrary at first, consider that the qualitative ZZ landscape depends mostly on the relative arrangement of the $\ket{\widetilde{11}}, \ket{\widetilde{02}},$ and $\ket{\widetilde{20}}$ states forming the two-photon manifold, which in turn depends on the qubit-qubit detuning $\Delta$ and on the anharmonicities $\alpha_{IST}$ and $\alpha_{Tr}$, but not e.g. on the absolute frequencies of the qubits. It is therefore only necessary to sweep two parameters to observe useful changes in $\Delta$ and in relative anharmonicities. Qualitatively similar results can be obtained by instead sweeping e.g. $E_{C,IST}$ and $E_{L,IST}$, or $E_{C,Tr}$ and $E_{J,Tr}$.

\begin{figure}[ht!]
\includegraphics[width=11cm]{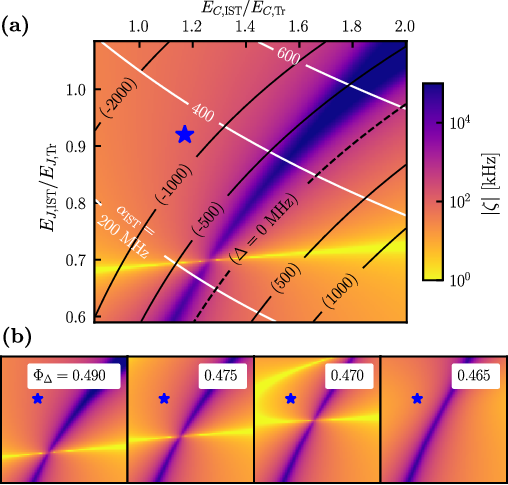} 
\caption{\label{fig:ZZ_simulations} Simulated landscape of static ZZ shift. (a) Magnitude of $\zeta$ versus circuit parameters $E_{J, \mathrm{IST}}$ and $E_{C, \mathrm{IST}}$ at $\Phi_\Delta=0.5$~$\Phi_0$, obtained from numerical diagonalization of the two-qubit Hamiltonian in Eq.~(\ref{eq:full_2QHam_ZZsim}) but with parasitic inductor modes excluded. All other quantities are fixed to the values extracted from device B, which is here denoted by the blue star. The black (white) lines indicate contours of constant IST-transmon detuning $\Delta$ (IST anharmonicity $\alpha_{\mathrm{IST}}$) calculated from the individual qubit spectra with $J=0$. (b) The same simulation is repeated while varying the flux bias. The $\zeta=0$ manifold sweeps across the parameter space, reaching device B close to $\Phi_\Delta=0.47$~$\Phi_0$, matching our experimental observation. Color scale and axes are the same as in (a).}
\end{figure}

 Compared to the simulation of flux dependence $\zeta(\Phi_\Delta)$ described above, we here perform an additional simplification by omitting the parasitic inductor modes from the IST Hamiltonian. This is to further reduce the simulation time, making it feasible to efficiently perform large two-dimensional scans of parameter space. Also, one can consider the parasitic modes as an extrinsic factor which could be removed in future design iterations, and we are therefore interested only in the intrinsic properties of an ideal IST-transmon system. Fig.~\ref{fig:ZZ_simulations}~(a) shows the resulting values of $\zeta$ as a function of the variables $E_{C,IST}$ and $E_{J,IST}$, written as ratios w.r.t. the fixed parameters $E_{C,Tr}$ and $E_{J,Tr}$. We choose to plot the absolute value $|\zeta|$ on a logarithmic scale to better visualize large changes in interaction strength. The main features are a quasi-horizontal region of weak ZZ interaction, and a diagonal region of maximal interaction strength. The former corresponds in fact to a one-dimensional manifold satisfying the condition $\zeta=0$, with $\zeta>0$ above it in this plot and $\zeta<0$ below. To better orient ourselves in parameter space, we add contours indicating constant values of $\Delta=\omega_{IST} - \omega_{Tr}$ (black lines) and of $\alpha_{IST}$ (white lines). Note how the $\zeta=0$ manifold crosses most of the $\Delta$ contours, which means that one can pick any desired detuning and always find a set of parameters that lead to suppressed ZZ interaction in an IST-transmon pair. This is crucial to maintain the freedom of frequency allocation in a larger multi-qubit processor. Additionally, the $\zeta=0$ manifold does not run parallel to the $\alpha_{IST}$ contours, further supporting the statement made in the main text that the naive condition $\alpha_{IST}=-\alpha_{Tr}$ is not very useful for predicting the optimal ZZ operation point in a real system.

 The region of maximal $\zeta$ corresponds to a large shift experienced by the $\ket{\widetilde{11}}$ computational state, i.e. when it is in close proximity to either the $\ket{\widetilde{02}}$ or $\ket{\widetilde{20}}$ states. A particularly interesting point is found when this region intersects the $\zeta=0$ line. This corresponds to the condition where $\ket{\widetilde{11}}, \ket{\widetilde{02}},$ and $\ket{\widetilde{20}}$ form a triple degeneracy point (equivalently, $\Delta=\alpha_{Tr}=-\alpha_{IST} \approx -220$~MHz in this case).

 It is interesting to observe how adding a flux bias in an experiment will modify the landscape shown in Fig.~\ref{fig:ZZ_simulations}~(a), which is fixed at the half-flux sweet spot $\Phi_\Delta=0.5$ of the IST. We repeat the same simulation for several values of $\Phi_\Delta$, shown by the sequence of panels in Fig.~\ref{fig:ZZ_simulations}~(b). Moving away from the flux sweet spot, the IST frequency increases while its anharmonicity decreases, causing the ZZ suppression condition to sweep across the parameter space. For any choice of circuit parameters such that $\zeta>0$ at the sweet spot (e.g. our experimental device B, denoted by the blue star), there is a particular value of  $\Phi_\Delta$ at which $\zeta$ goes through zero. This does not happen if one starts from a condition where $\alpha_{IST}$ is already too small at the flux sweet spot to fully balance out the opposite effect of the (negative) transmon anharmonicity.  The presence of a ZZ-suppressed operation point is therefore guaranteed across a wide variety of device parameters, as long as applying a flux bias is acceptable in practice. On the other hand, it is generally desirable to simultaneously achieve low ZZ error and maintain high coherence, and therefore target a system with $\zeta=0$ as close as possible to $\Phi_\Delta=0.5$ in order to avoid increased decoherence due to flux noise. The insights gained from these numerical simulations provide guidelines for future device design.

\section{\label{sec:app_fluxnoise} Flux noise characterization}

An important limiting factor for a flux-tunable circuit is its sensitivity to magnetic flux noise. While operation at a flux sweet spot can protect from most of the dephasing coming from this noise channel, it is often necessary to flux-tune a qubit away from this point, such as in the case of the sideband interactions between a transmon and IST explored in this work. We therefore perform a measurement of dephasing versus applied external flux on the IST of device B to characterize the flux noise spectrum that the circuit is exposed to, using the method described in Ref.~\cite{yoshihara_decoherence_2006}. 

\begin{figure}[t]
\includegraphics[width=9.5cm]{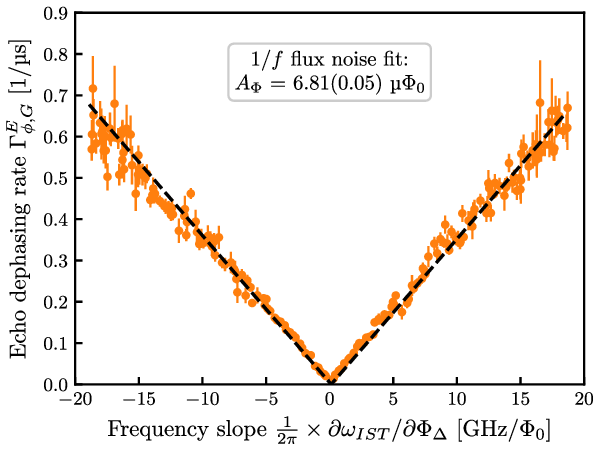} 
\caption{\label{fig:IST_deviceB_fluxnoise} Determination of $1/f$ flux noise amplitude for IST qubit on device B. An echoed coherence measurement is performed across a wide range of IST external flux $\Phi_\Delta$, and the resulting decay trace is fitted to a Gaussian envelope to extract the pure dephasing rate $\Gamma_{\phi, G}^E$. This quantity is proportional to the slope of the frequency flux-dispersion curve, $\partial\omega_{IST}/\partial \Phi_\Delta$. A linear fit to this data directly provides the underlying $1/f$ flux noise amplitude $A_\phi$.}
\end{figure}

SQUIDs are typically subjected to $1/f$-type flux noise, with a power spectral density (PSD) $S_\Phi(\omega)=A_\phi^2/|\omega|$, where $A_\phi$ is the characteristic flux noise amplitude scale. While the exact source of this noise source is still under investigation, experimental and theoretical results point to a microscopic origin of flux noise, likely due to local magnetic impurities fluctuating and creating a noise background that couples to the SQUID loop, typically on the order of $A_\phi \sim 1-10$~\unit{\micro\Phi_0}~\cite{yoshihara_decoherence_2006, koch_model_2007, kumar_origin_2016, braumuller_characterizing_2020}. The noise amplitude $A_\phi$ can be extracted from the dependence of the qubit's pure dephasing rate $\Gamma_\phi$ on flux. When dephasing is limited by $1/f$ noise, the decoherence curve has a Gaussian envelope rather than the typical exponential envelope for broadband noise. We choose to perform a spin-echo sequence to provide long enough coherence times for easier fitting, even under the presence of large flux noise, and we take into account the (more or less constant) exponential decay contribution from relaxation $T_1$ and flux-independent decoherence $T_2^E$, which we set to the values measured at the sweet spot and reported in Table~\ref{tab:device_overview}. This allows us to extract the purely Gaussian contribution $\Gamma_{\phi,G}^E$ to the echoed dephasing rate, which is plotted in Fig.~\ref{fig:IST_deviceB_fluxnoise}. The flux noise amplitude can then be extracted by fitting the simple linear relation~\cite{yoshihara_decoherence_2006, braumuller_characterizing_2020}
$$
\Gamma_{\phi,G}^E = \sqrt{\ln(2)}A_\phi \times \left| \frac{\partial\omega_{IST}}{\partial \Phi_\Delta}\right|,
$$
where the frequency-flux dispersion $\partial\omega_{IST}/\partial \Phi_\Delta$ is obtained from the numerical fit to the IST spectroscopy data in Fig.~\ref{fig:IST_deviceB_spectroscopy}. The linear fit agrees well with the data, from which we obtain $A_\phi = 6.81(0.05)$~\unit{\micro\Phi_0}. This is comparable to typical values measured in our gradiometric transmons under similar conditions~\cite{Unpublishedfluxpaper}, suggesting that the IST design does not suffer from significantly increased flux noise. 

This result is encouraging but also somewhat surprising, given that there is evidence in the literature that a SQUID loop with a higher aspect ratio (longer and thinner wires making up the perimeter of the loop) should be exposed to higher total flux noise amplitudes, all else being equal~\citep{bialczak_1f_2007, braumuller_characterizing_2020, peruzzo_geometric_2021}. One explanation is that we might currently be limited mostly by external flux noise, which could be mitigated through improved shielding of the device and filtering of the flux lines. Beyond this, we believe that the extremely large aspect ratios achievable in an RF-SQUID qubit with a geometric inductor, such as our IST design or the large superinductance qubits of Ref.~\citep{peruzzo_geometric_2021}, could open a new avenue for the study of flux noise in a much wider parameter range than achieved with DC-SQUIDs in the past. This could in turn lead to new insights into its microscopic nature and to a potential strategy for its mitigation.

\providecommand{\noopsort}[1]{}\providecommand{\singleletter}[1]{#1}%